\documentclass[prl,twocolumn,graphicx,amssymb,floatfix]{revtex4}

\usepackage{graphicx}
\begin{document}

\title{Paradoxes of the Aharonov-Bohm and the Aharonov-Casher effects. }

\author{L. Vaidman}
\affiliation{ Raymond and Beverly Sackler School of Physics and Astronomy\\
 Tel-Aviv University, Tel-Aviv 69978, Israel}

\begin{abstract}
 For a believer in  locality of Nature, the Aharonov-Bohm effect and the Aharonov-Casher effect are  paradoxes. I discuss these and other Aharonov's paradoxes and propose a local explanation of these effects.  If the solenoid in the Aharonov-Bohm effect is treated quantum mechanically, the effect can be explained  via local interaction between the field of the electron and the solenoid. I  argue that the core of the Aharonov-Bohm and the Aharonov-Casher effects is that of quantum entanglement: the quantum wave function describes all systems together.
    \end{abstract}
\maketitle

Thirty years ago, after calculating integrals for the QCD sum rules in the M.Sc. studies at Weitzmann Institute I went to Tel Aviv looking for a more intuitive physics research. There I met  Yakir Aharonov who suggested to look at his recent papers on nonlocal measurements \cite{AA}. I found  a very different approach. Assume we can measure a certain nonlocal variable, say a product of two variables related to spatially separated  regions. Then we can send signals faster than light: a paradox! A paradox that led us to find measurable nonlocal variables, define new types of measurements, construct a new formalism of nonlocal measurements \cite{AAV86,VNL}.

For Yakir, paradoxes are the main tool for developing new physics.  Yakir is certain that we will have a new revolution in physics and the paradoxes we  find  lead us toward it. During last thirty years I learned from Yakir how to use paradoxes to do research. However, our philosophical approaches became different.  I want to believe that apart from some (important) details we understand Nature today. Yakir taught me to use paradoxes as a powerful research tool, but instead of leading to new directions, it leads me to correct, improve, deepen and clarify our current physical theories.

At the beginning of the twentieth century in spite of serious paradoxes of physical theories the view that physics is ``finished'' was shared by many, but the theory of relativity and the quantum theory proved them all
 wrong. Today, when physics explain almost everything we can see, physicists rarely claim that today's physics is close to the final theory of the Universe.  Quantum theory brings two elements which make it very difficult to believe that we completely understand Nature: randomness and nonlocality. Yakir accepts randomness and nonlocality: ``God does plays dice to avoid contradiction with nonlocality''. For me, accepting randomness is accepting limits to physics.   I refuse to accept it. I have to pay a big price for this: the only consistent way  to avoid randomness in outcomes of quantum experiments that I can see is to accept that all outcomes take place in Nature. Thus, I have to accept existence of numerous parallel worlds corresponding to all possible outcomes of quantum experiments and adopt the many-worlds interpretation of quantum mechanics (MWI)\cite{Everett,SEP}.

The MWI avoids, together with randomness, the nonlocality of the Bell-type correlations. When we have entangled particles, measurement of one of them changes nothing in local description of the other. It was a mixed state before the measurement and remains the same mixture after it. In the picture of the whole Universe  which incorporates all the worlds measurement of an entangled particle causes no action at a distance.  In a particular world, created by a quantum measurement, we experience an illusion of randomness and can observe nonlocal  correlations.

There is one type of nonlocality which the MWI does not remove, the nonlocality of the Aharonov-Bohm (AB) effect \cite{AB}. The AB effect does not lead to ``action at a distance'', but it prevents a local explanation of the dynamics of charged particles in particular setups.  This nonlocality is what I want  to analyze here.

In classical physics we can explain the behavior of particles in the following local way: Particles create fields (propagating with velocity of light or slower) and the other particles accelerate due to local action of these fields.  Wave packets of classical electromagnetic field (which are not waves of some media made of particles) also change their propagation due to local interaction. The interference of overlapping electromagnetic waves packets is a local phenomenon too.

Consider a  Mach-Zehnder interferometer (MZI), Fig. 1a. If the interferometer is properly tuned,  the wave packets split inside, but invariably reunite toward detector $A$. Small shift of one of the mirrors increasing the length of one arm by a half a wave length leads to  a classical lag between the wave packets  which changes the interference after the final beam splitter such that the wave packet ends at detector $B$, Fig. 1b. All this behaviour is  perfectly understood by local interaction of the wave packets with beam splitters and mirrors and finally local interference of the wave packets moving toward detectors.

\begin{figure}[b]
  \includegraphics[height=6.0cm]{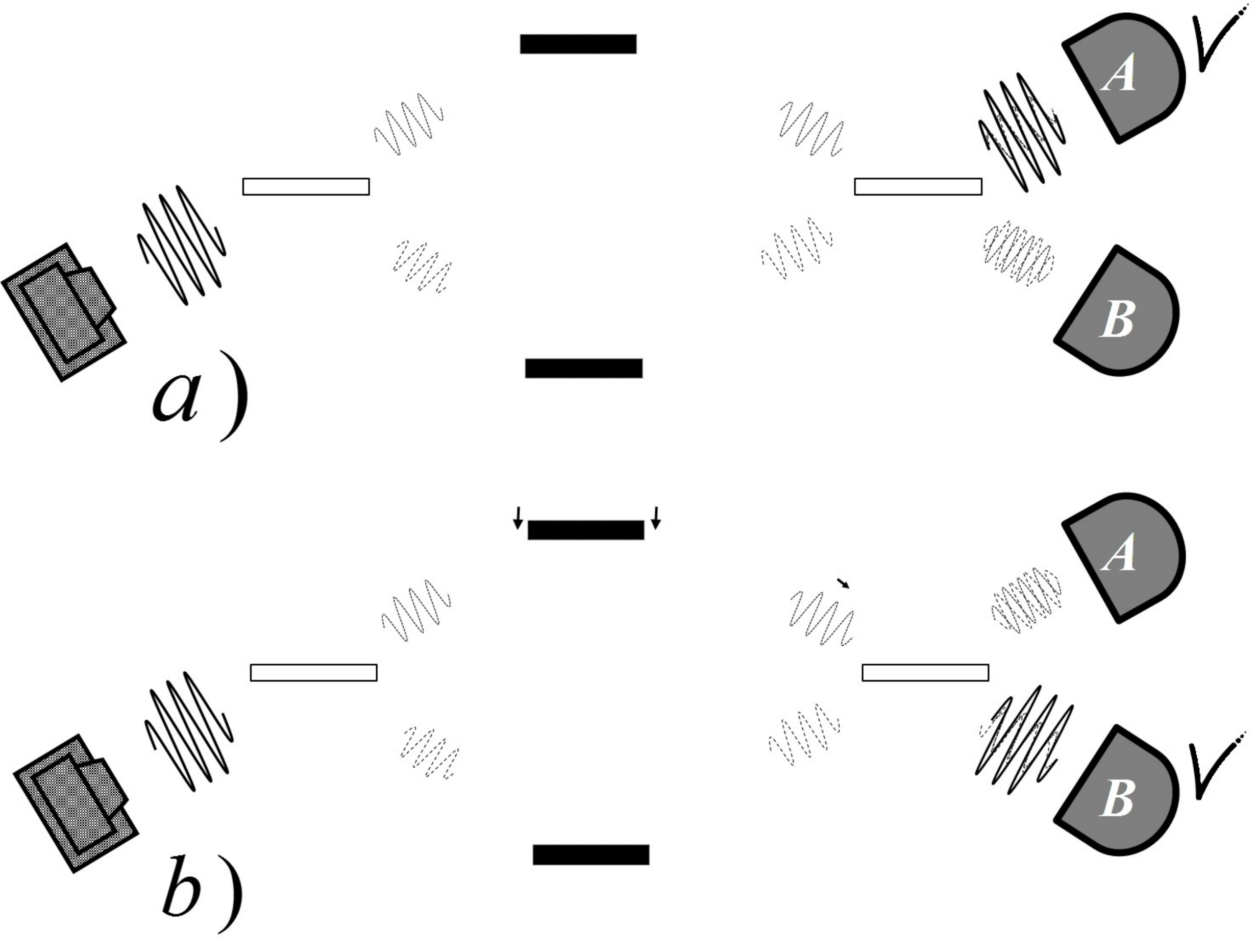}\\
      \caption{{\bf Mach-Zehnder interferometer.}  The wave packet entering the interferometer splits in two beam splitters creating two wave packets moving toward detector $A$ and another two moving toward detector $B$. a). The MZI is tuned in such a way that the pair of wave packets moving toward $A$ interfere constructively and the pair moving toward $B$ interfere destructively. b). Moving a little the upper mirror shifts the upper wave packet which causes shifts of one of the wave packet moving toward $A$ and one moving toward $B$. The shift by the half of the wavelength results in destructive interference toward $A$ and constructive interference toward $B$.   } \label{1}
\end{figure}

\begin{figure}[b]
  \includegraphics[height=2.6cm]{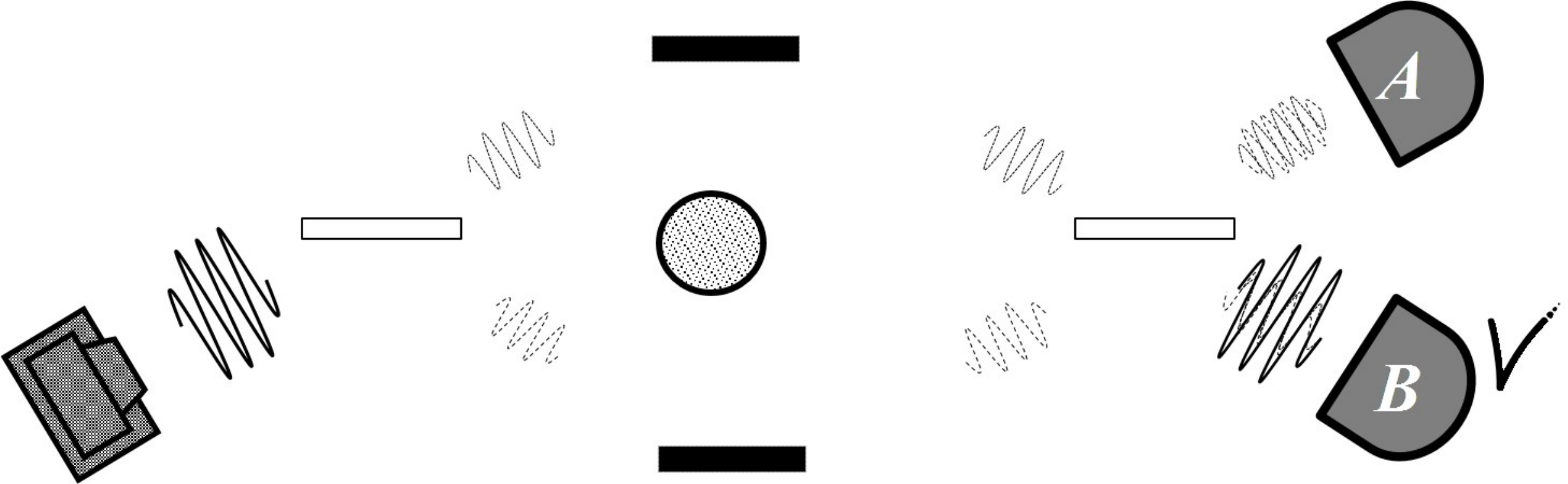}\\
      \caption{{\bf The Aharonov-Bohm effect.}  The electron MZI with   a solenoid inside the interferometer  exhibits the AB effect. The wave packets are not shifted, but the solenoid leads to a relative phase between the lower and the upper wave packets which causes the change in the interference picture from constructive to destructive interference toward detector $A$, and destructive to constructive interference toward detector $B$.   } \label{2}
\end{figure}

More than thirty years ago the interference experiment with MZI has been performed with single photons and, not surprisingly, showed the same results \cite{Gra}. It is natural to assume that it can be explained in the same way as above with the replacement of a classical wave by the quantum wave of the photon. However, while the evolution of the wave packets of classical electromagnetic fields in the MZI setup can be observed locally, the evolution of the quantum wave is unobservable. The relative phase of the wave packets of the quantum wave which controls the interference when they finally overlap cannot be observed locally when the wave packets are still separated. (There is a possibility of measuring  the relative phase of the wave packets using local coupling to parts of  a composite  measuring device \cite{AVNL}, but then the result depends on the definition of the relative phase of the measuring device, so it does not provide an unambiguous measurement of the relative phase of the photon's wave packets.) Although the phase of each wave packet could not be measured during the propagation of the packet inside the interferometer, until the discovery of the Aharonov-Bohm effect \cite{AB}, there was no reason to suspect that it changes in a conceptually different way than the phase of a classical wave packet.

To observe the AB effect, we introduce a solenoid with a flux $\Phi$ of the magnetic field  inside an electron  MZI, Fig. 2. The interferometer is tuned in such a way that when the flux vanishes, the electron ends at detector $A$ with certainty. The solenoid leads to a relative phase $\phi_{AB}=\frac{  e \Phi}{c\hbar}$ and choosing the flux such that $\phi_{AB}=\pi$  causes the electron to change the interference and to end at detector $B$.

Contrary to the case where we moved a mirror and thus locally changed the evolution of a wave packet in one of the arms, we do not have a local explanation of the change in the evolution of the electron wave packets in the present case: The electromagnetic field of the solenoid vanishes at the trajectories of the wave packets of the electron and the vector potential can be made to vanish at any point along the trajectories by using the gauge freedom. The line integral of the vector potential which equals  the enclosed magnetic flux is  gauge invariant and proportional to the AB phase. However, this provides a topological rather than a local explanation as we had before.

 In the following, I will present an attempt to provide a local explanation of the AB effect. However,  first I discuss the Aharonov-Casher  (AC) effect \cite{AC} which is dual to the AB effect, and show that  Boyer's local explanation \cite{Boyer} of the AC effect fails.  Aharonov and Casher noticed that there is a symmetry in the interaction between a polarized neutron and an electron. A vertical line of vertically polarized neutrons is equivalent to a solenoid of the AB effect. The symmetry of the interaction suggests that the replacement  electron $\leftrightarrow$ neutron for all particles will transform the AB effect in the electron MZI with a solenoid to an analogous  AC effect in the neutron MZI with a line of charges. In the AC effect, in contrast to the AB effect, the neutron wave packets do not move in the field-free region. The magnetic field, however, is zero, so naively the neutron does not experience an electromagnetic force, but Boyer  correctly realized that a commonly used current-loop model of the neutron leads to a non-vanishing electric dipole moment for a moving neutron and thus it experiences the electromagnetic force due to the electric field of the line of charges. The neutron accelerates while approaching the line of charges in one arm of the interferometer and decelerates in  the other arm, then decelerates (accelerates) to the original velocity until it leaves the vicinity of the charged line. The classical lag between  the two wave packets provides a local explanation of the AC effect.

 Boyer's paper was published when Yakir, I,  and Philip Pearle were together in South Carolina.  Aharonov's intuition was that Boyer cannot be right. He came with the following paradox: Let us consider elastic mirrors for the neutron approaching the charged line, see Fig. 3. Since the induced electric dipole moment $\vec{d}=\frac{\vec{V}\times\vec{\mu}}{c}$ changes its direction together with the change of the direction of the velocity, the Boyer's force $(\vec{d} \cdot\vec{\triangledown})\vec{E}$  always accelerate in our setup. But nothing else is changed in the system, so the Boyer's force is a free source of energy!

\begin{figure}[b]
  \includegraphics[height=3.6cm]{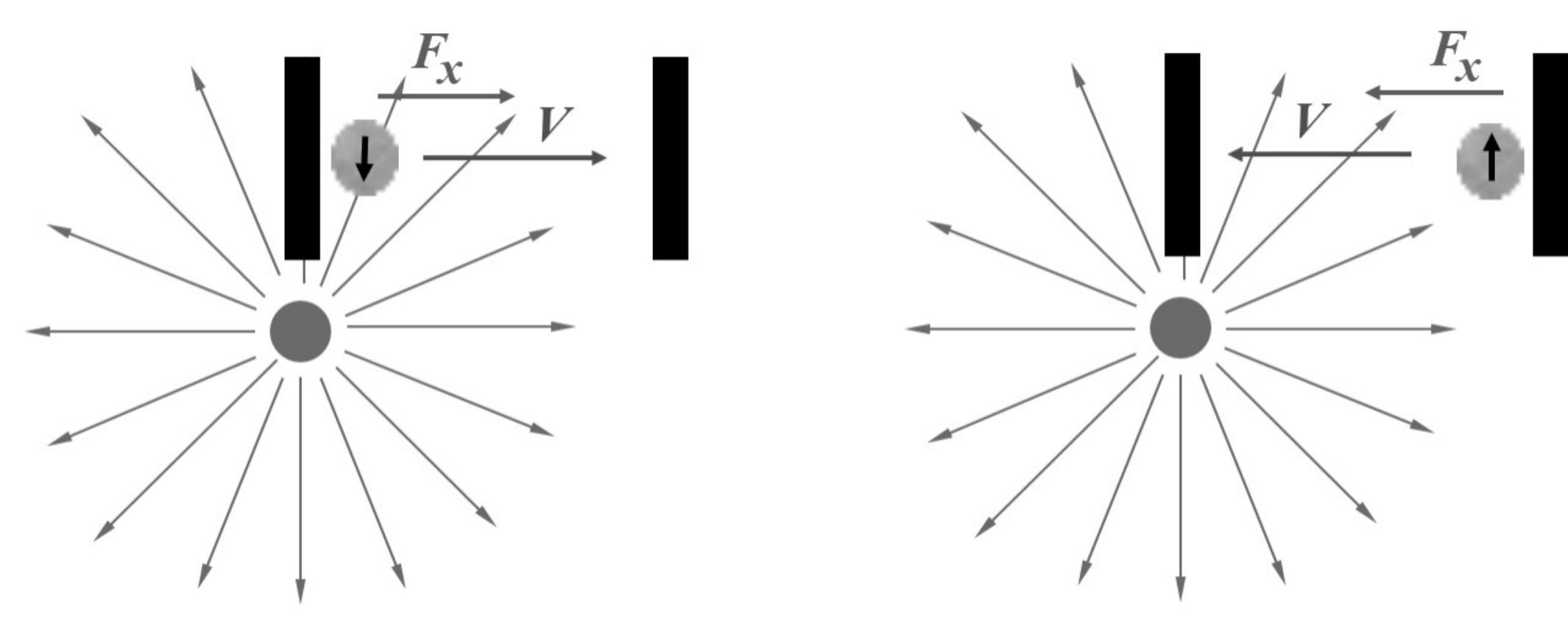}\\
      \caption{{\bf Free energy source.}  According to Boyer, a polarized neutron bouncing between two mirrors in the presence of the line of charge will experience force in the direction of its motion. Thus it can be used as a never ending source of energy.    } \label{3}
\end{figure}

 After some time we found the resolution of the paradox \cite{APV,torq}. Boyer provides the correct expression for the force, but in this case the Second Law of Newton is more subtle than just $\vec{F}=m\vec{a}$. A current loop in an electric field has a ``hidden'' mechanical momentum $\frac{\vec{\mu}\times\vec{E}}{c}$ and the Boyer's force  just provides the time derivative of this momentum. There is no acceleration, so (unfortunately) there is no free source  of energy. There is  no classical lag between the wave packets and, therefore, the AC effect exhibits the same paradoxical nonlocal feature as the AB effect.  In fact, later we learned that this hidden momentum is the core of the paradox discussed by Shockley and James \cite{SJ} almost a half century ago. (Note  that  apparently the hidden momentum is not well enough understood until today, see the erroneous conclusion published just a few months ago \cite{Mans}.)

 So, we  have now two effects which have no local explanation. The wave packets of the electron move in an identical way with or without the solenoid, or for   the neutron, with or without the charged line but, nevertheless, the interference depends on the electromagnetic sources. It seems that there is only global explanation of these effects.  The final interference depends on the integral on a closed trajectory and it is apparently meaningless to ask in which part of the trajectory the influence of the solenoid (line of charges) took place.

For me this is a paradox. At every place on the paths of the wave packets of the particle there is no observable effect of any kind, but nevertheless,  a relative phase is generated. In an attempt to find local explanation for everything I can see, I identify  a weak point in the current descriptions of the AB and the AC effects: the fields with  which the quantum particles interact are considered to be classical. I believe that everything is quantum. And indeed, considering the solenoid as a quantum object I have found a local explanation for the AB effect \cite{VAB}.
 I will show that the AB effect arises from  different  shifts of the wave packets of the source of the magnetic flux which experiences different local electric fields created by the two  wave packets of the electron  inside the MZI.

Consider the following setup. The solenoid consists of two  cylinders of  radius $r$, mass $M$, large length $L$  and  charges $Q$ and $-Q$ homogenously  spread  on their surfaces. The cylinders   rotate  in  opposite directions with surface velocity $v$. The electron encircles the solenoid with velocity $u$ in superposition of being in  the left and in the right sides of  the circular trajectory  of radius $R$, see Fig.~4.

\begin{figure}[b]
  \includegraphics[width=6.5cm]{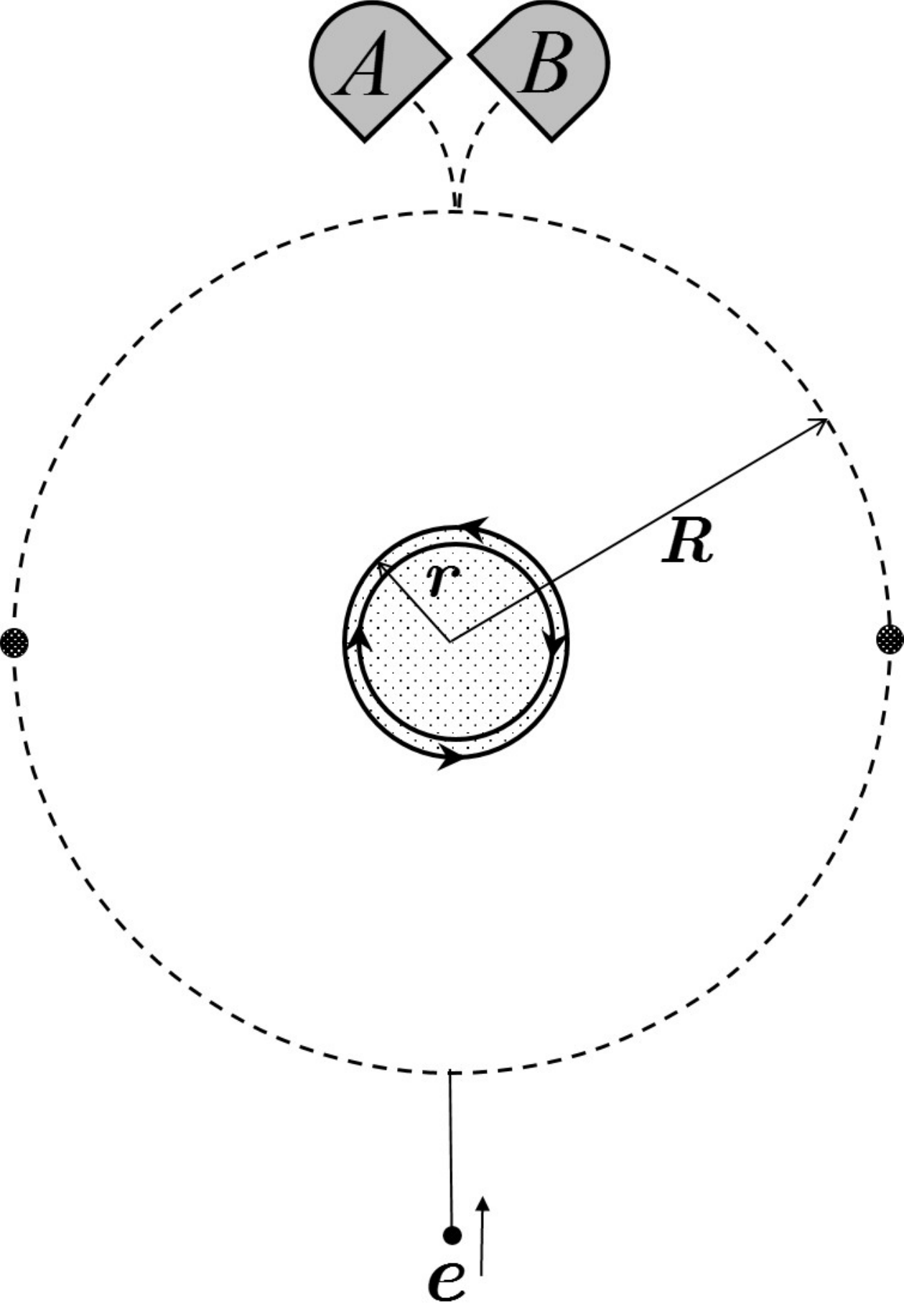}\\
      \caption{{\bf Solvabe model of the magnetic AB experiment.} The electron wave packet coming directly toward a circle splits into a superposition of two wave packets and after encircling the solenoid in the center interfere on the beam splitter toward detectors $A$ and $B$. } \label{4}
\end{figure}

The flux  in the solenoid due to two cylinders is $2 ~\pi r^2 ~ \frac{4\pi}{c} \frac{Qv}{2\pi r L}  =\frac{4\pi Q v r}{cL}$, and the AB phase, i.e., the change in the relative phase between left and right wave packets due to electromagnetic interaction is:
\begin{equation}\label{ABpha}
\phi_{AB}=\frac{4\pi e Q v r}{c^2L\hbar}.
\end{equation}
When the electron moves on the circular trajectory, it creates magnetic flux through a cross section of the solenoid   seen at angle $\theta$, see Fig.5,
\begin{equation}\label{phitheta}
\Phi(\theta)=\pi r^2  \frac{u e \cos \theta}{c(\frac{R}{\cos \theta})^2}=\frac{\pi r^2 e u \cos^3 \theta}{c R^2}.
\end{equation}
 Before the electron entered the circle, it provided no flux through this section. By entering one arm of the circle, the electron produces change in the the magnetic flux and  causes an electromotive force on charged solenoids  which change their velocity. The change in the velocity is:
\begin{equation}\label{deltav}
\delta v= \frac{1}{M}  \int \frac{\pi r^2 e u\cos^3 \theta}{c^2 R^2}\frac{1}{2\pi r} \frac{R }{\cos^2 \theta} 2\pi r \frac{Q}{2\pi r L} d\theta=\frac{uQer}{c^2MRL}.
\end{equation}
The shift of the wave packet of the cylinders due to this velocity change during the motion of the electron wave packet is
\begin{equation}\label{deltax}
 \delta x=\delta v \frac{\pi R}{u}=\frac{\pi Qer}{c^2ML}.
\end{equation}
We consider here motion on the circle of radius $r$ as a linear motion. The relevant wavelength of de Broglie wave of each cylinder is $\lambda= \frac{h}{Mv}$. For calculating the AB phase we should take into account that both cylinders are shifted and that they shifted (in opposite directions) in both branches. This leads to factor 4 and provides correct expression for the AB phase:
 \begin{equation}\label{deltax}
 4\frac{2\pi \delta x}{\lambda}=\frac{8\pi^2 Qver}{hc^2L}=\phi_{AB}.
\end{equation}

\begin{figure}[b]
  \includegraphics[width=7.2cm]{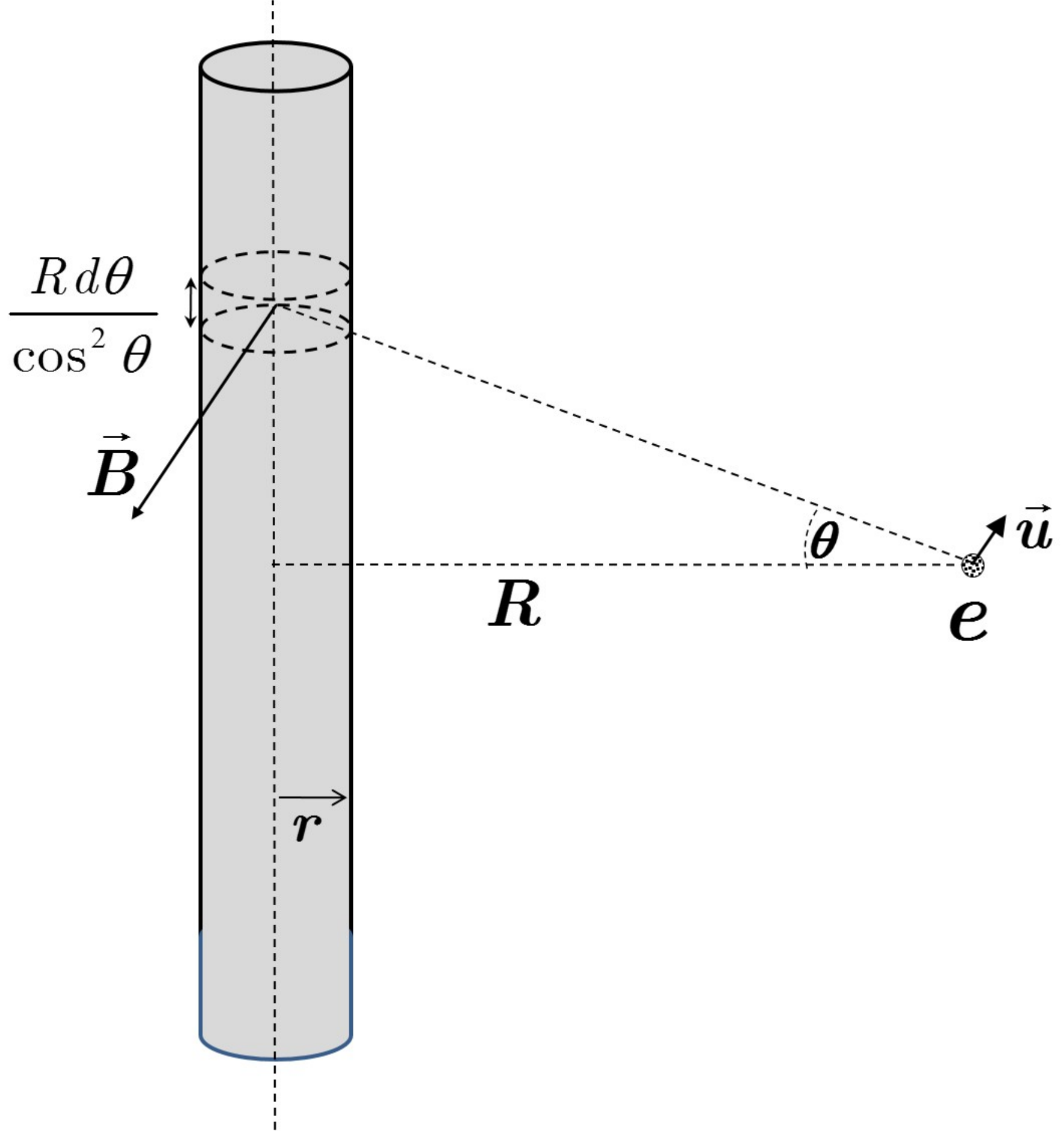}\\
      \caption{{\bf Magnetic field created by the electron.}   The field at the cross section of the cylinder when the right wave packet is seen at angle $\theta$. } \label{5}
\end{figure}

The explanation of the AB effect is as follows. The electron in superposition of two arms of the interferometer creates different fields  at the location of the  source of the potential and thus the wave packet of the source is shifted differently. The change in the wave function of the source is, essentially, the change in the relative phase only. In the AB effect, due to topology of the circle, the envelope of the wave packet has no any change whatsoever. Since in quantum mechanics the wave function is for all parts of the system together, the change in the wave function of the source leads to observable effect in the interference experiment with the electron.

Another manifestation of the wholeness of quantum mechanical description is entanglement. In fact, during the process of observation of the AB effect, the shift in the momentum of the source might create entanglement with the electron; it  disappears  before the end of the experiment. But this entanglement is not necessary: the initial uncertainty in momentum of the source might be much larger than the intermediate shift in the momentum due to the interaction with the electron.   Thus, although for the explanation it is crucial to understand that the quantum wave function is for all particles, the effect can appear even if the source particles and the electron at all times are described well by a product state.

The pictorial explanation with spatial wave packets shifted by fields will disappear when  we go beyond physics of moving charges. A model for a solenoid can be a line of polarized neutrons. If the model of each neutron is a current loop, then we will have the picture as above, but if we describe it as a quantum spin, we will not have a picture of charged particles kicked by the field of the electron, we should say that the magnetic field of the electron  changes the phase of the neutrons directly. (Note, however, that viewing the polarized spin state of the neutron as a superposition of two spin states polarized in perpendicular direction, we can restore the story of electron fields causing rotation around the axes of the solenoid.) This is also an explanation of the AC effect: the local field acting on the neutron is responsible for appearance of the AC phase.

One of the most revolutionary aspects of the AB effect is that contrary to classical mechanics, in quantum mechanics we cannot  explain everything by action of local fields. Potentials, which were auxiliary objects in classical physics, have direct physical meaning in quantum mechanics.  Explanation of the AB and the  AC effects in local terms allows us to entertain the idea that potentials are auxiliary concepts after all. However, today all versions of quantum theory, and the Schr\"{o}dinger equation in particular, are based on the concept of potential. There is no quantum analog of the Second Law of Newton. This work might open the challenge of developing a theory which will tell us how  quantum particles  evolve when the interaction between them  is described by  local fields which have no potential.

Even  before developing of such a local theory, my assertion provides one useful corollary: If the fields vanish at locations of all particles then these fields yield no observable effect.   In the magnetic AB effect I see no variant for which all systems move in  a free-field area, but I can devise such an example for the electric AB effect. Consider the following configuration of three charged particles on a straight line: In the center we place an electron with charge $-e$. On  both sides at equal distances we place two charges $4e$. Immediate calculation shows that electric field created by any two particles at the location of the third particle vanishes. The electric potential   at location of the electron due to other charges does not vanish, but according to my corollary it cannot cause any effect. The effect of potential might naively be expected when we consider an electron MZI in which we bring the two charges to the electron wave packet moving in one arm of the interferometer keeping all the time the configuration described above, see details in reference \cite{VAB}. The subtlety here is that we have to bring the charges toward one arm of the interferometer only in the ``branch'' in which the electron wave packet is there. The charges are in the mixed quantum state of being near and far from the interferometer. The charges do not provide classical field and this explains why the standard approach to the AB effect fails.

Classical mechanics has local formulation with Newton's laws and global formulation  via Lagrangian and Hamiltonian. Quantum mechanics has only global formulations: Hamiltonian, Feynman's path integrals etc.  In the past I also learned  from the AB effect that the potentials have direct observable effects in quantum mechanics and thus quantum mechanics cannot have a local formulation \cite{HE}. But the local explanation of the AB and the AC effects presented above give me hope for a local quantum mechanics.   I know that Yakir's intuition is against local theory I am looking for (see, however,  a discussion of similar ideas in Aharonov's recent publication \cite{AKN09}). As a reply to my proposal, Yakir, as usual, presented me a challenge in the form of a paradox: ``Consider everything quantum, as you do, and explain in your local terms the AB effect experiment by Tonomura \cite{Tonomura} in which the solenoid was screened by a superconductor and the AB phase (of the value of $\pi$) was observed in a very convincing way.'' The superconductor apparently screens the field of the electron, so my proposed mechanism for the AB phase via the motion of the charges in the solenoid fails. I guess that the phase appears due to a local action of the electron on charges in the superconductor, but I still cannot provide an  explanation. Whatever the resolution of this paradox is,  it will deepen our understanding of quantum mechanics as many other Aharonov's paradoxes already did  in a very profound way.

I thank  Shmuel Nussinov for useful discussions. This work has been supported in part by the Binational Science Foundation Grant No. 32/08 and
 the Israel Science Foundation  Grant No. 1125/10,


\begin{thebibliography}{99}

\bibitem{AA}
Y. Aharonov and D. Albert, Phys. Rev. D {\bf 21}, 3316 (1980); ibid. {\bf 24}, 359 (1981).

\bibitem{AAV86}
Y. Aharonov, D. Albert and L. Vaidman,
Phys. Rev. D {\bf 34}, 1805 (1986).

\bibitem{VNL}
L. Vaidman
Phys. Rev. Lett. {\bf 90}, 010402 (2003).


\bibitem{Everett}
H. Everett III,
 Rev. Mod. Phys. {\bf 29}, 454  (1957).

\bibitem{SEP} L.~Vaidman,  Many-Worlds Interpretation of Quantum
Mechanics, {\it Stan. Enc. Phil.},  E. N. Zalta (ed.) (2002),
http://plato.stanford.edu/entries/qm-manyworlds/.

\bibitem{AB}
Y. Aharonov and D. Bohm,
 Phys. Rev. {\bf 115}, 485 (1959).

\bibitem{Gra}
 P. Grangier, G. Roger and A. Aspect,  Europhys. Lett. {\bf 1} 173 (1986).

\bibitem{AVNL}
Y. Aharonov and L. Vaidman
Phys. Rev. A {\bf 61}, 052108 (2000)

\bibitem{AC}
Y. Aharonov and A. Casher,
Phys. Rev. Let {\bf 53},   319   (1984).

\bibitem{Boyer}
T. Boyer,
  Phys. Rev. A {\bf 36},    5083   (1987).


\bibitem{APV}
Y. Aharonov, P. Pearle,   and L. Vaidman,
Phys. Rev. A {\bf 37},    4052  (1988).

\bibitem{torq}
L. Vaidman,  Am. J. Phys. {\bf 58}, 978 (1990).

\bibitem{SJ}
W. Shockley and R. P. James,  Phys. Rev. Lett. {\bf 18}, 876 (1967).


\bibitem{Mans}
M. Mansuripur,  Phys. Rev. Lett. {\bf 108}, 193901 (2012).

\bibitem{VAB}
 L. Vaidman
Phys. Rev. A {\bf 86}, 040101 (2012).

 \bibitem{HE}
 L. Vaidman, Aharonov-Bohm Effect,  in {\it Hebrew Encyclopedia}, supplementary volume III (in Hebrew) (1994).


 \bibitem{AKN09}
Y. Aharonov, T. Kaufherr, and S. Nussiov,
        J. Phys.: Conf. Ser. {\bf 173},  012020 (2009).


 \bibitem{Tonomura}
A. Tonomura {\it  et al.} Phys. Rev. Let {\bf 56},   792   (1986).




\end{thebibliography}
\end{document}